# Surface electron perturbations and the collective behavior of atoms adsorbed on a cylinder

Boris Dzyubenko[1], Hao-Chun Lee[1], Oscar E. Vilches[1], and David H. Cobden[1]*

[1]Department of Physics, University of Washington, Seattle WA 98195, USA

*Corresponding author: cobden@uw.edu

**A single-walled carbon nanotube presents a seamless cylindrical graphene surface and is thus an ideal adsorption substrate for investigating the physics of atoms and molecules in two dimensions and approaching the one-dimensional limit[1-7]. When a suspended nanotube is made into a transistor, frequency shifts of its mechanical resonances allow precise measurement of the adsorbed mass down to the single-atom level[8-10]. Here we show that its electrical characteristics are also modified by the adsorbed atoms and molecules, partly as a result of a small charge transfer between them and the carbon surface. We quantify this charge transfer, finding it similar for many different species, and use the associated sensitivity of the conductance to carry out studies of phase transitions, critical scaling, dynamical fluctuations, and dissipative metastable states in a system of interacting atoms confined to a cylindrical geometry.**

Graphite is the substrate of choice for investigating the collective two-dimensional (2D) behavior of adsorbed atoms and molecules, including transitions between 2D solid (S), liquid (L) and vapor (V) phases[11, 12]. In addition, the physical interactions of adsorbates with carbon are important in filters, electrodes, sensors and gas storage, but very little is known experimentally about their effects on the electrons in the surface. Using the unique combination of assets of suspended nanotube transistors, held in equilibrium vapor[6], we are able to detect for the first time the charge transfer from neutral atoms or molecules. Surprisingly, it is of a similar magnitude for all the simple gases tested ($^4$He, Ar, Kr, Xe, $N_2$, CO and $O_2$), and although small (much less than predicted[13-15]), at gate voltages near threshold it can produce a large change in conductance. Thus, simply by monitoring the conductance we are able to explore the phase transitions of atoms on a cylinder, seeing 2D critical and triple points and critical behavior matching the 2D Ising universality class with a finite-size cutoff. We also observe intriguing new features in the phase transition dynamics, and discover nonlinear effects of adsorbates interacting with electrical current.

Each device, containing a nanotube of diameter ~2 nm and suspended length ~1 μm, is mounted in a vapor cell at temperature $T$ and pressure $P$, the latter being deduced from the pressure $P_g$ on an external gauge as indicated in Fig. 1a (see also Supplementary Information S1). The nanotubes have small band gaps, producing a minimum in the conductance $G$ near zero gate voltage[16] $V_g$, as can be seen in the characteristics of a device (YB11) shown in Fig. 1b. At lower temperatures, contact barriers cause the nanotube to act as a single-electron transistor exhibiting reproducible Coulomb blockade (CB) oscillations, visible in the characteristic at 4.3 K (blue).

Using the arrangement indicated in Fig. 1a, while measuring the conductance we can also detect mechanical resonances and deduce the coverage, $\phi$, which is the number of adsorbates per carbon atom (see Methods). For each gas, as the pressure is increased we find that a single dense monolayer forms before the vapor saturates. During this process the $G - V_g$ characteristics change slightly: for example, Fig. 1b shows characteristics of YB11 at 47 K in vacuum (black) and in saturated argon (red). The change in $G$ is generally a few percent, and varies between devices (see Supplementary Information S2 for more examples), but it always has one common feature: a distinct shift of the "gap" along the $V_g$ axis, by an amount $\Delta V_g$. This shift is usually most apparent and unambiguous at the p-channel threshold, which tends to be sharper. However, in the case of helium we can see that the CB peaks shift roughly uniformly over the entire gate voltage range, as illustrated in Fig. 1c for another device, YB14. Such behavior is not consistent with a change in capacitance, or in the contact transparency, but rather implies that a net voltage-independent charge is donated by the adsorbates to the nanotube, presumably associated with the hybridization of the adsorbate's orbitals with the surface electron states. From the period $\Delta V_{CB}$ of the CB oscillations, which corresponds to adding the charge $-e$ of one electron, we can directly convert $\Delta V_g$ to a change $\Delta Q = +e\Delta V_g/\Delta V_{CB}$ in the total charge on the nanotube.

Fig. 1d shows how $\Delta V_g$ changes with argon coverage on device YB11 at several temperatures. In the limit of dense coverage ($\phi \sim 0.15$), $\Delta V_g$ approaches $\sim +10$ mV, independent of $T$. With $\Delta V_{CB} = 11$ mV this gives $\Delta Q \approx +0.9e$, amounting to an average charge transfer per Ar atom of $\Delta Q/N_A \sim +2 \times 10^{-5}e$, where $N_A = \phi N_C \sim 5 \times 10^4$ is the number of argon atoms and $N_C \sim 2.5 \times 10^5$ the number of carbon atoms in the suspended part of the nanotube. In the limit of sparse coverage, however, $\Delta V_g$ is approximately proportional to $\phi$, and we can obtain an average charge donated per atom from the slope. It is negative and grows in magnitude with decreasing $T$, reaching about $-7 \times 10^{-4}e$ at 42 K. Both limits are plotted together in Fig. 1e. This behavior is fairly typical, but there are large variations between devices in the sparse coverage limit. For example, with device YB14 there is no apparent negative swing and the CB peak shift is monotonic in helium pressure, as shown in Fig. 1f, though slightly different for the p-channel and n-channel peaks.

Close to threshold the change in $G$ at fixed gate voltage can be large, allowing it to be measured much more precisely than $\Delta V_g$. In Fig. 2a we plot both $G$ and $\phi$ at a series of equilibrium pressures at 47 K for device YB30. Importantly, the upward swing in $G$ matches the riser in $\phi$, implying that most of the change in $G$ is a result of adsorption on the vibrating part of the nanotube, whatever the mechanism. The decrease in $G$ preceding the riser also tracks the variation of $\phi$, but with the opposite sign. The inset shows $G$ vs $\phi$, which is approximately linear at low density (red line) as expected if in this limit the contribution of each argon atom is independent.

The variation of $G$ with $P$ at fixed $T$ can be measured conveniently by gradually increasing the pressure, at the expense of some accuracy in $P$. Fig. 2b compares "conductance isotherms" obtained in this way for Ar, CO and N$_2$ at 50 K on device YB27, and Fig. 2c compares Ar and O$_2$ at different temperatures on device YB24. We see that the behavior of all these gases is strikingly similar, implying that their hybridization with the nanotube electrons is also very similar. Although this might at first seem surprising, it is consistent with the known fact that they have similar physisorption binding energies on graphite[12]. All are compact molecules/atoms of similar mass with stable electronic configurations and low polarizability, and our results suggest that the physics of the binding is quantitatively generic to such cases.

The effect varies between gases, but with the following common features (see Supplementary Information S2). First, at dense coverage ($\phi \approx 0.15$) there is a charge transfer to the nanotube of the order of $e$ in total, or around $10^{-5}e$ per molecule, which is almost independent of temperature. For Ar, $\Delta Q$ ranged from $+0.25e$ to $+0.9e$; for N$_2$, from $-0.4e$ to $+0.2e$. Second, the charge transfer (as determined by CB shifts) is almost independent of gate voltage, and thus of a large electric field approaching 1 V/nm normal to the nanotube surface. Third, the change in conductance away from the gap on forming a monolayer is usually only a few percent, implying that the backscattering of electrons from adsorbates is very weak. Last, at sparse coverage the charge transfer is often of the opposite sign, grows as $T$ decreases, and varies in magnitude, from undetectable (in device YB14) to about $-10^{-3}e$ per molecule (in YB11). One factor in explaining this may be that while some nanotubes, such as YB14, can be deduced to be extremely clean from the phase transitions discussed below, others such as YB11 may have patches of contamination or amorphous carbon. The first adsorbates will attach to higher-binding sites surrounding the patches, where they might also induce a different charge. However, the amount of contamination must be small enough not to produce detectable features in the low-pressure portion of the coverage isotherms. In any case it is clear that the large responses to simple gases previously reported for nanotube[17-20] and graphene[21, 22] devices were not associated with the clean surface but with contacts, substrate, contamination, or defects[23, 24], and our results provide a new improved basis for theoretical understanding of adsorption on carbon substrates.

We can now exploit the conductance as a powerful probe of the adsorbate system. It can be sensitive to changes in ordering as well as density; it does not rely on locating a vibrational resonance; and it can be measured with high precision in linear response, so minimally perturbing the system. For the device held near threshold in Fig. 2a, changing $N_A$ by one at small $\phi$ corresponds to a measurable fractional change in $G$ of ~$10^{-3}$ (although we would expect $N_A$ to fluctuate in equilibrium on a microsecond time scale). This sensitivity permits investigation of the collective behavior of 2D matter on a cylinder. Recent simulations have predicted commensuration effects on the phases on the cylindrical nanotube surface[4], but for simplicity we focus here on argon, which is less likely to exhibit commensurate solid phases.

For Ar on graphite, the 2D triple point and critical points[25, 26] are $T_{tr} \approx 48$ K and $T_c \approx 55$ K, both lower than the 3D equivalents, with uncertainty of about $\pm 0.5$ K in each (see Supplementary Information S3). Some nanotube devices, including YB14, show features matching these, implying that Ar on a clean nanotube surface forms a homogeneous quasi-2D system in which the interactions between atoms are the same as on graphite. Fig. 3a shows conductance isotherms for Ar on YB14. The main riser becomes vertical below about 55 K (red traces), as expected for a first-order V-L transition emerging below $T_c$, with continuous supercritical behavior at higher temperatures (black traces). A small additional riser (green arrows), which is easier to identify in very slow pressure sweeps (upper inset) and disappears below 48 K (blue traces), indicates an L-S transition occurring above $T_{tr}$. The pressures at the risers are shown on a $P - T$ diagram in Fig. 3b. They are higher than the pressures at the phase boundaries on graphite, indicated by dashed lines. Part of the difference is a systematic error of up to a factor of two caused by the cell pressure $P$ lagging the gauge pressure $P_g$. The lower inset to Fig. 3a illustrates the hysteresis due to this lag at 51 K, and how it can be reduced by slowing the sweep rate to home in on the true equilibrium pressure at the L-V jump, which is plotted as an open red circle in Fig. 3b. The remaining order-of-magnitude pressure excess is due to weaker binding to nanotubes than to graphite[1].

The conductance can also be employed to investigate critical behavior. According to the theory of critical phenomena[27] the compressibility of the monolayer, proportional[11] to $(\phi^2 T)^{-1} d\phi/d(\ln P)$, should diverge approaching $T_c$ in the supercritical regime as a negative power of the reduced temperature, $\tau \equiv (T - T_c)/T_c$. This behavior cannot be studied in coverage measurements because the vibrational resonance is too distorted in this regime, but since $G$ is a smooth function of $\phi$ we anticipate the same scaling in the conductance isotherms. Fig. 3c shows the maximum slope, $dG/d(\ln P)$, of each isotherm vs $T$. At 55 K and below, no matter how slowly we introduce gas, there is an instantaneous jump and the measurement is truncated by the 100 ms circuit response time. Inset is a log-log plot of $(1/T)dG/d(\ln P)$ vs $\tau$ taking $T_c = 55.5$ K from the literature[26]. For larger $\tau$ it follows the power law $\tau^{-\gamma}$ where $\gamma = 7/4$, the critical exponent for the 2D Ising model universality class which we expect to apply here. On the other hand the points at $\tau < 0.03$ lie below the extrapolated power law, consistent with finite-size suppression of the power-law divergence[28] roughly when $1/\tau$ exceeds the nanotube circumference divided by the diameter of an argon atom. This is the first experimental system to allow studies of critical behavior in 2D with a periodic boundary condition.

The conductance also reveals the dynamics of the phase transitions. Fig. 4a shows a series of closely spaced conductance isotherms near $T_c$. Growing temporal fluctuations are seen on approaching $T_c$, where critical fluctuations are expected. They cannot be caused by sample temperature variations, which would also produce fluctuations away from $T_c$. In addition, below $T_c$ a sharp dip appears at the foot of the V-L step. A similar anomaly appears below $T_{tr}$ at the top of the V-S step (bold arrow), as can be seen in Fig. 3a, while none is seen at the top of the V-L step or on either side of the L-S step. These anomalies have large fluctuating components, as illustrated in Fig. 4b where we monitor $G$ while nudging the pressure repeatedly up and down through the L-V transition at 51 K. In both directions the anomaly appears on the V side but not on the L side of the transition. The number of fluctuations recorded depends on the precise pressure sweep rate. A closer look at a transition with many fluctuations is shown below.

Instead of using the conductance as a passive probe, we can pass a current large enough to perturb the adsorbates and produce nonlinear effects. Fig. 5a shows a set of I-V sweeps, both up and down, measured in several pressures of Ar vapor at 47 K. They exhibit a reproducible kink, with negative differential resistance, that moves to higher bias as $P$ increases. The kink is a transition between a higher conductance level, obtained when a monolayer is present, and a lower conductance level matching that obtained in vacuum (black). When the conductance $I/V$ is plotted against power $VI$ (Fig. 5b) each kink becomes a nearly vertical jump, and another smaller jump is visible at lower power. The jumps can be attributed to phase transitions resulting from Joule heating of the nanotube above the cell temperature. Assuming the nanotube temperature $T_{tube}$ is uniform (justified because the heat is generated mainly at the contacts) and its rise is proportional to the power, we expect $T_{tube} = T + \alpha VI$ where $\alpha$ is a constant. By requiring consistency between measurements at different cell temperatures we derive $\alpha = 0.088$ K/nW. In Fig. 5c we plot the jump positions in terms of $T_{tube}$ at each pressure. The small jumps (green circles) and the larger ones (red circles) lie on curves close to the S-L and L-V boundaries, respectively, in Fig. 3b. We conclude that I-V measurements can be used as isobaric temperature sweeps for rapidly mapping the phase diagram.

We see a variety of other nonlinear phenomena. For example, Fig. 5d shows I-V sweeps at 80 K in Xe vapor ($T_{tr} = 99$ K), which exhibit sharp jumps. On cycling $V$ there is hysteresis and metastability (Fig. 5e), a natural consequence of positive feedback because in this case the conductance, and thus heating, increases when the monolayer evaporates. However, the current is not a single-valued function of power (Fig. 5f), and the middle of the three levels has additional structure and a large difference in current from the low-bias S state, hard to reconcile with a simple uniform equilibrium L state. There are also anomalies near the jumps reminiscent of those seen in the conductance isotherms, and other unexplained features including the sub-steps at $I/V \approx 8$ μS in the large jumps in Fig. 5b. Further studies using other adsorbates and nanotubes of known chirality will be needed to reveal the nature of such phenomena and to build a full understanding of coupling between adsorbates and conduction electrons.

**Methods:**

The nanotubes are grown by chemical vapor deposition using $H_2/CH_4$ at 800 °C and $Fe(NO_3)_3/MoO_2$ catalyst, across trenches between pre-patterned Pt source and drain electrodes[16] on $Si_3N_4/SiO_2$. The trenches are 1 μm wide and 0.5 μm deep, with a Pt gate on the trench bottom. We select nanotubes with small gaps (20-100 meV) because they have better electrical contacts than large-gap ones. The vapor cell is mounted in a closed-cycle cryostat with temperature range 4 K – 300 K. The gas handling manifold and ceramic capacitance pressure gauge are at room temperature. The cell pressure $P$ is lower than the gauge pressure $P_g$ in static equilibrium due to thermal transpiration and is inferred by applying a correction[29]. The linear-response conductance $G$ is measured using a 1 mV ac source bias at 600 Hz, a virtual-earth current preamplifier and a lock-in amplifier. The nonlinear I-V is measured using a dc source bias. The coverage $\phi$ is measured by adding to the source bias a swept radio-frequency signal at frequency $f$ amplitude-modulated at $f_{am} = 1$ kHz (see Supplementary Information). Mechanical resonances[30] are identified as peaks in the mixing current measured by a lock-in amplifier referenced to $f_{am}$. The equation $\phi = N_a/N_c = (m_C/m_a)\left[(f_0/f_P)^2 - 1\right]$ gives a good measure of the coverage $\phi$, where $f_P$ and $f_0$ are the positions of the mechanical resonance at pressure $P$ and in vacuum respectively, $m_a$ is the molecular mass of the adsorbate, and $m_C$ is the atomic mass of carbon[6].

**Acknowledgments:**

This work was supported by NSF DMR award 1206208. Silicon structures were fabricated in the UCSB Nanofabrication Facility and the UW NTUF. We thank Richard Roy for contributions in setting up the experiment, and Marcel den Nijs for helpful discussions.

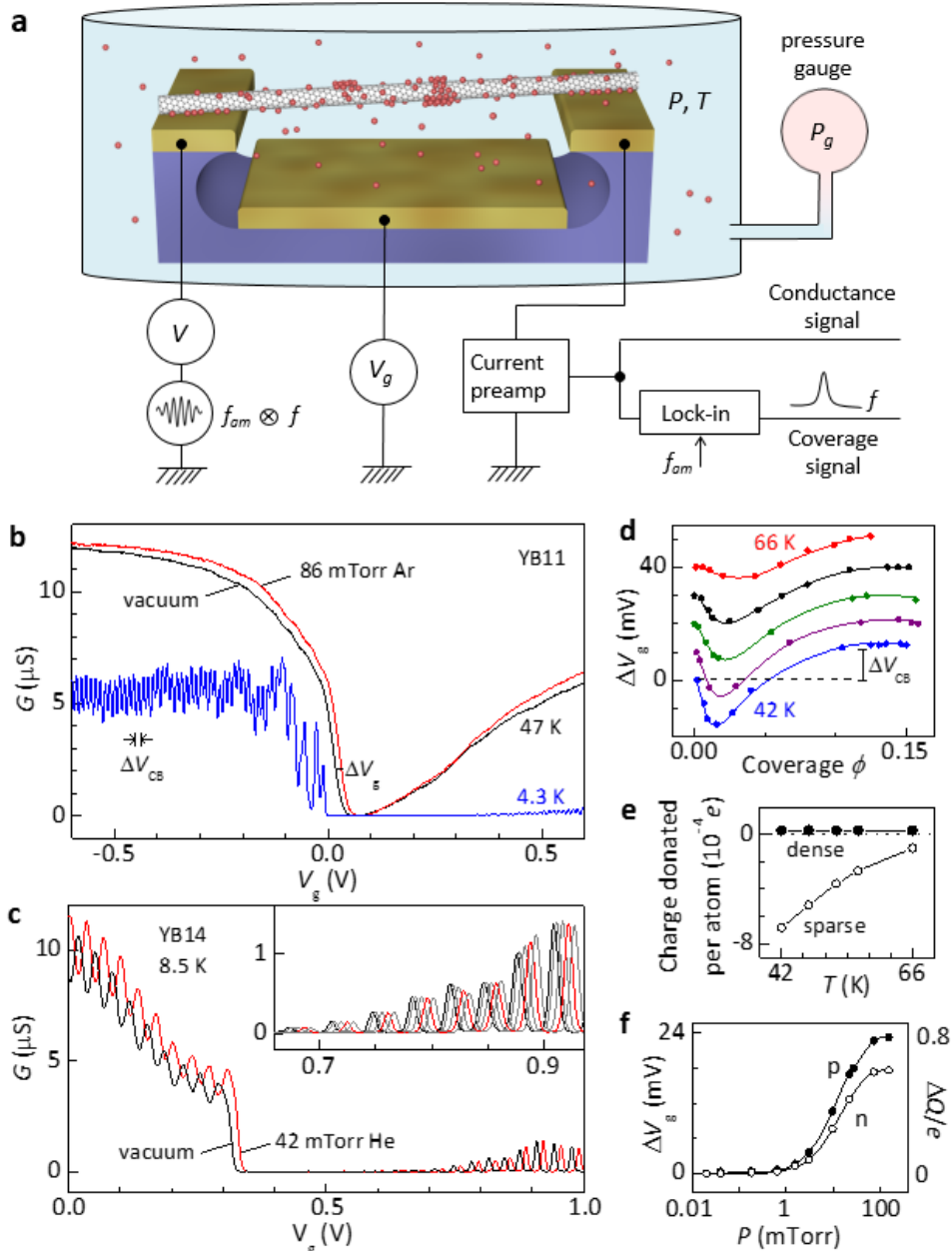

**Figure 1 | Effects of adsorbates on nanotube conductance. a**, Schematic arrangement for measuring both the conductance $G$ and the coverage $\phi$ (using a vibrational frequency shift) of a suspended nanotube in a vapor cell at temperature $T$ and pressure $P$ connected to an external gauge reading pressure $P_g$. **b,** Characteristics of device YB11: in vacuum (black) and coated with a monolayer of Ar (red, $P_g$ =86 mTorr) at 47 K, showing a threshold shift $\Delta V_g$; and in vacuum at 4.3 K (blue) showing Coulomb blockade oscillations with period $\Delta V_{CB}$. **c,** Effect of $^4$He on device YB14 at 8.5 K. Inset: traces at 0.1 (black), 8, 21, 42 (red), and 220 mTorr, showing a steady, reproducible shift. **d,** Threshold shift vs coverage at temperatures $T$ = 42, 47, 52, 56, and 66 K (offset sequentially by 10 mV for clarity). Smooth curves are guides for the eye. **e,** Approximate charge transferred to the nanotube per Ar atom vs $T$ at sparse (open circles) and dense (solid circles) monolayer coverage.. **f**. Shifts vs $^4$He pressure for the p-channel (solid) and n-channel (open circles) peaks in **c**.

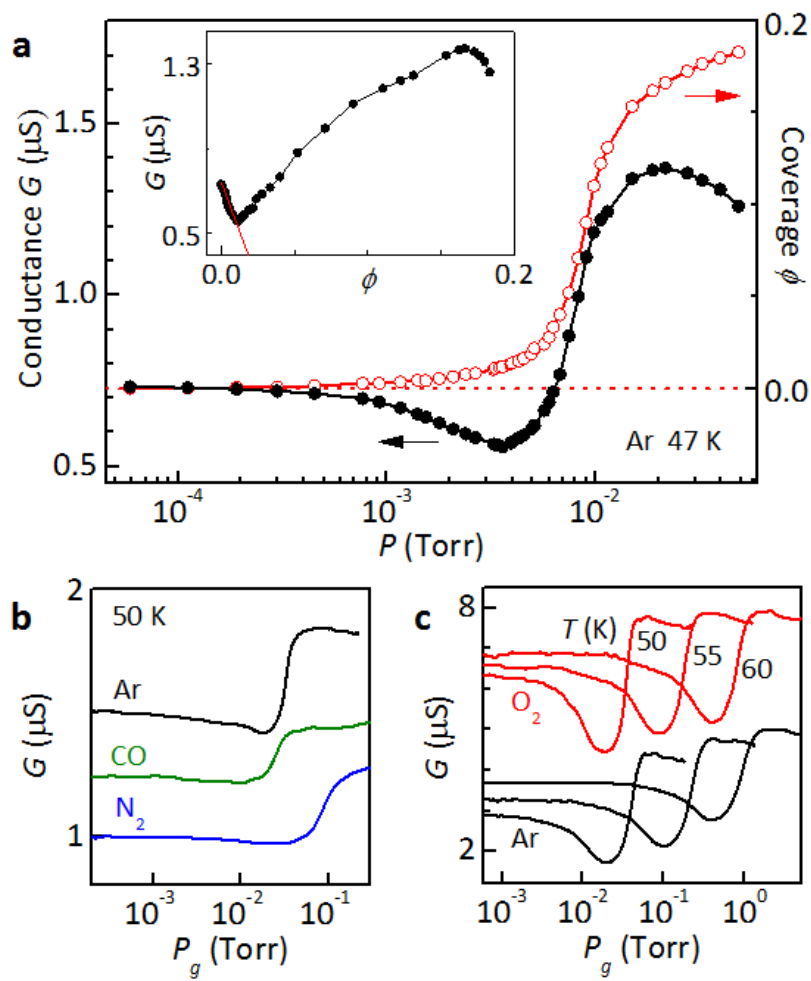

**Figure 2 | Conductance isotherms. a**, Coverage $\phi$ (open red circles, right axis) and conductance $G$ (solid black circles, left axis) measured together at a series of equilibrium Ar pressures at 47 K (device YB30, $V_g = -0.05$ V). Inset: $G$ vs $\phi$, with low-density linear region indicated by a red line. **b**, Continuous conductance isotherms at 50 K for Ar, CO and $N_2$ (device YB27; $V_g = -1.0$ V; upper traces offset for clarity). **c**, Comparison of Ar and $O_2$ conductance isotherms at several temperatures (device YB24; $V_g = -0.5$ V; $O_2$ traces offset for clarity).

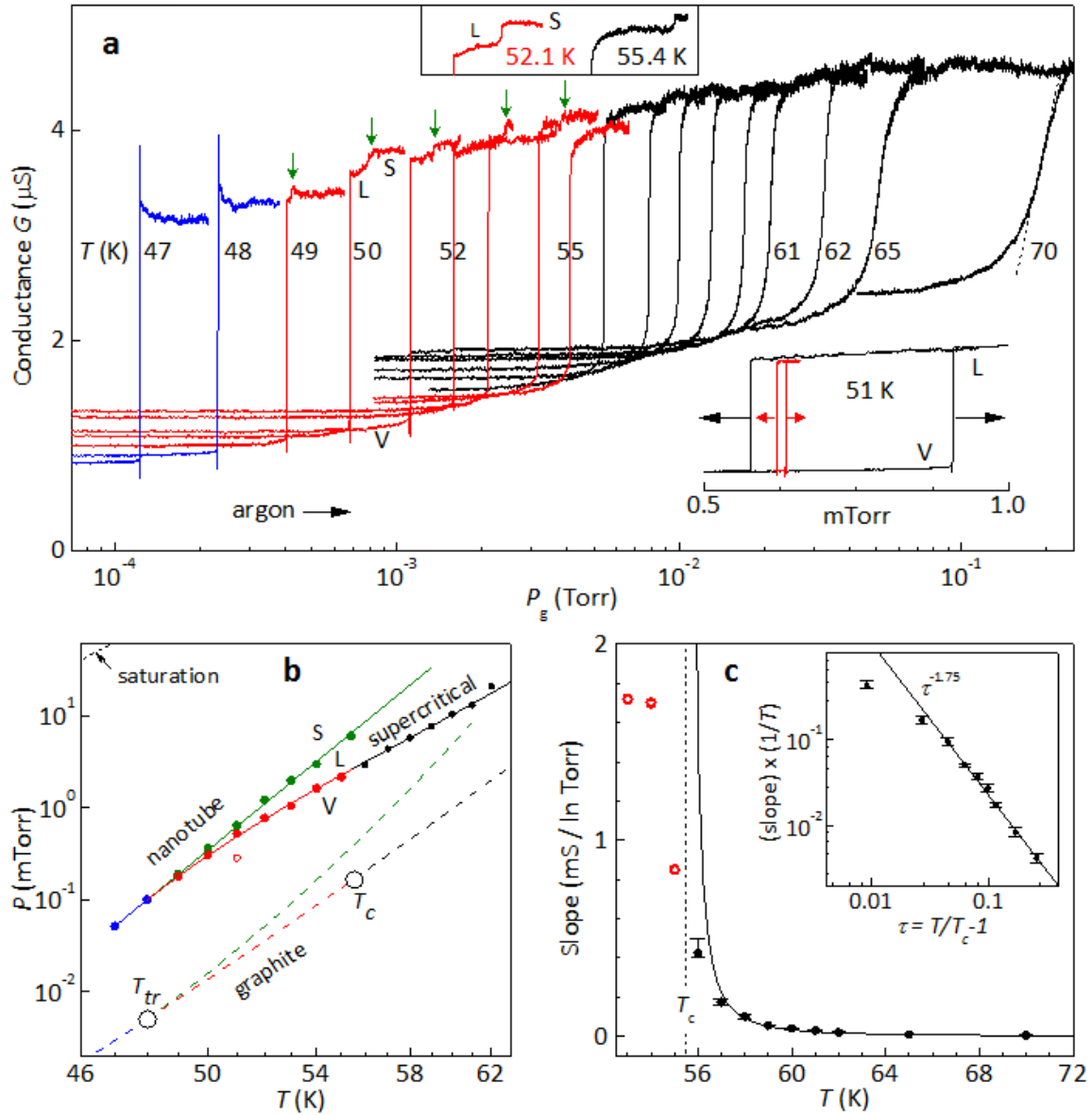

**Figure 3 | Probing the phases boundaries and critical behavior of two-dimensional argon. a**, Series of conductance isotherms each taken with Ar pressure increasing (device YB14, $V_g = 0.34$ V). The interval is 1 K below 62 K. Blue traces are below the inferred triple point, $T_{tr} \approx 48$ K; red traces are between $T_{tr}$ and the critical point, $T_c \approx 55.5$ K; and black traces are supercritical. Upper inset: sections of slower isotherms showing the smaller riser (indicated by green arrows) more clearly. Lower inset: hysteresis caused by the cell pressure lag at 20 mTorr/hr (black) and 1 mTorr/hr (purple). **b,** Phase boundaries between 2D vapor (V), liquid (L) and solid (S) Ar on this nanotube deduced from the measurements in (a). Red and blue points are the positions of the larger riser, and green points of the small riser. The extra point (open circle) at 51 K has no gauge lag error. Dashed lines are the corresponding 2D phase boundaries on graphite. **c**, Maximum slope of the main riser vs temperature. Error bars show range of straight line fits consistent with the data. Measurements below $T_c$ (red open circles) are limited by the system response. Inset: log-log plot of slope ($\times 1/T$) vs reduced temperature $\tau$. The line indicates the inverse power $\gamma = 1.75$ for the 2D Ising universality class.

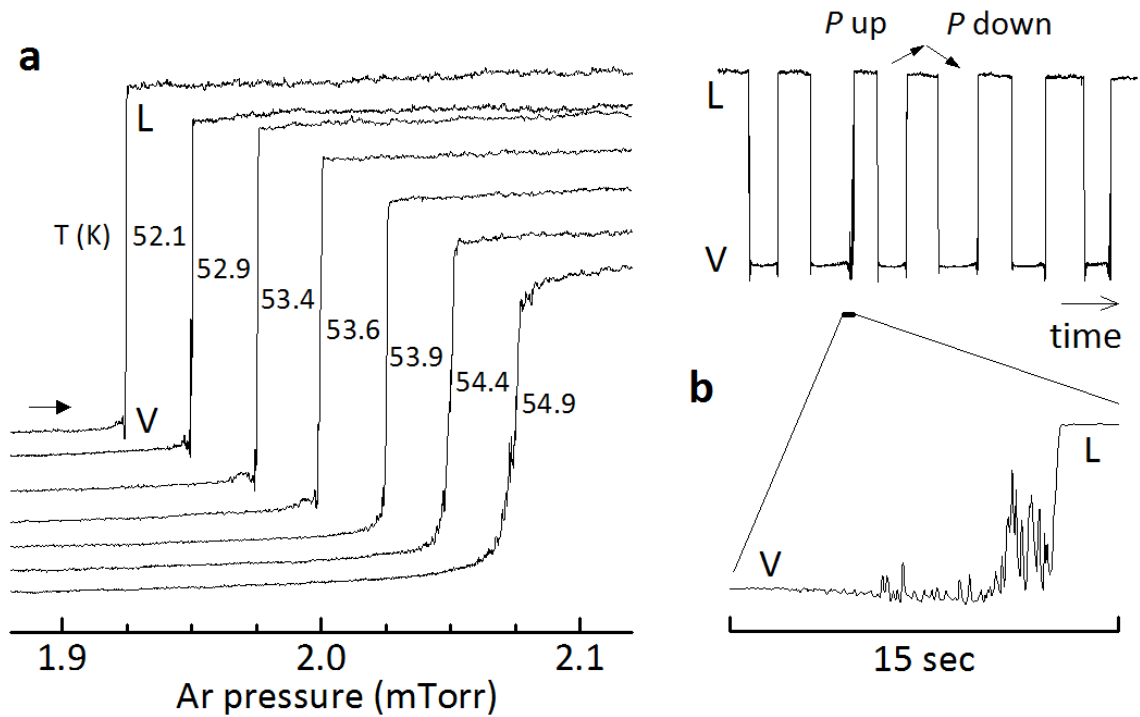

**Figure 4 | Dynamical effects at a 2D phase transition. a**, Conductance isotherms close to the 2D critical point, $T_c$ = 55.5 K, for Ar on device YB14. Traces are individually offset along both axes for clarity; the pressure is increased at ~0.16 mTorr/min; and the circuit response time is 100 ms. As the riser becomes vertical, a sharp dip with associated temporal fluctuations appears at its foot. **b,** Conductance monitored while gently increasing and decreasing the pressure at 51 K to repeatedly cross the V-L transition. Below is a zoom on one V→L jump.

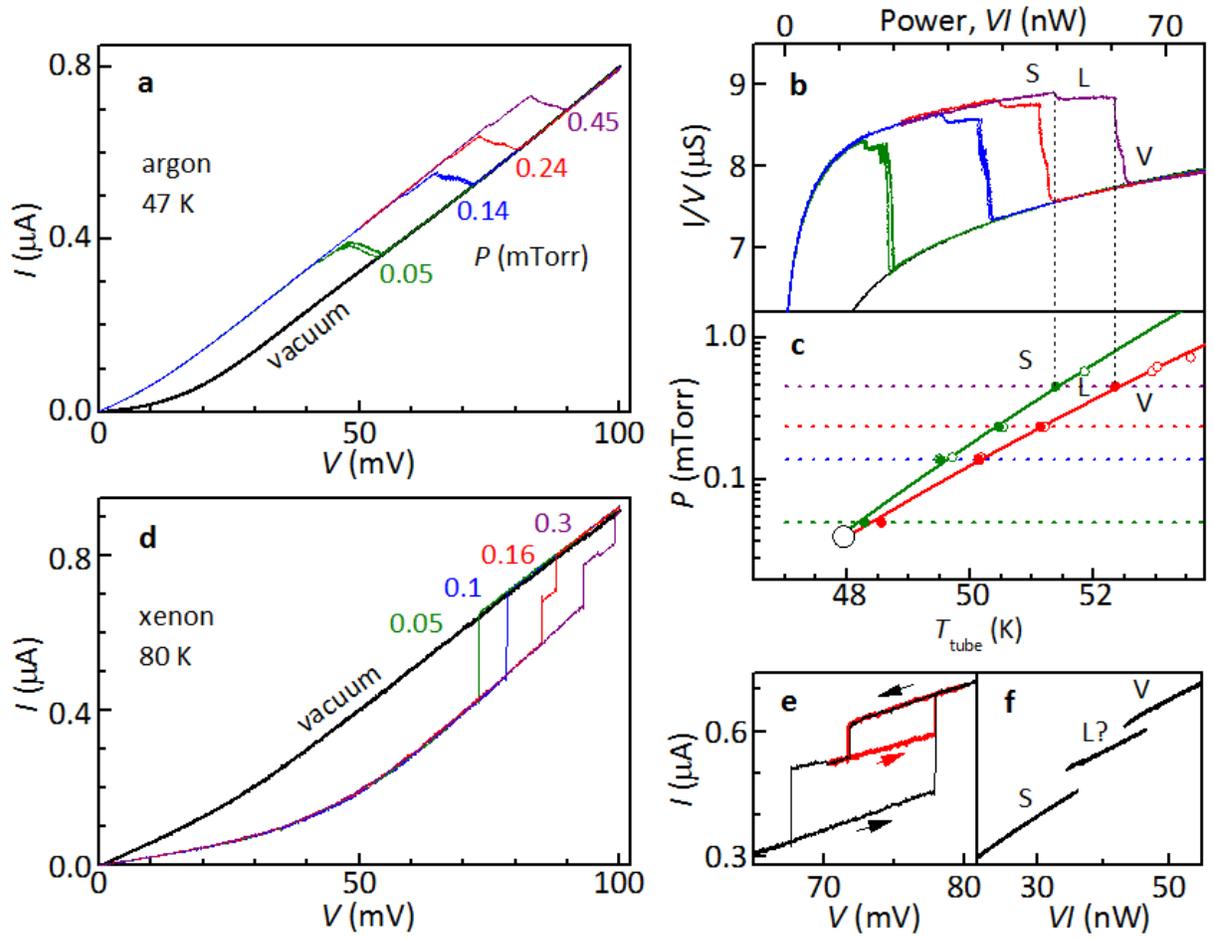

**Figure 5 | Nonlinear behavior. a**, I-V measurements on device YB14 in Ar vapor for several pressures at cell temperature 47 K ($V_g = 0.34$ V). Both up and down sweeps are plotted in each case. **b**, The same data plotted as conductance ($I/V$) vs power ($VI$), showing near-vertical jumps linked to phase transitions in the Ar monolayer. **c**, The power at each jump is mapped to nanotube temperature (bottom axis; see text) and the positions of smaller steps are plotted as green solid circles, the larger steps as red solid circles. Open circles are from measurements at higher cell temperatures. The results lie close to the phase boundaries obtained from isotherms in Fig. 3. **d**. I-V measurements in Xe vapor at 80 K (here sweeping $V$ up only). **e.** Cycling $V$ reveals multiple metastable current levels ($P_g = 0.16$ mTorr). **f.** Same data plotted vs power, indicating S, V and presumed L states.

# SUPPLEMENTARY MATERIAL
# Surface electron perturbations and the collective behavior of atoms on a cylinder

Boris Dzyubenko[1], Hao-Chun Lee[1], Oscar E. Vilches[1], and David H. Cobden[1]*

[1]Department of Physics, University of Washington, Seattle WA 98195, USA

*Corresponding author: cobden@uw.edu

## S1. Experimental details

*Devices*: The nanotubes are grown by chemical vapor deposition using $H_2/CH_4$ at 800 °C and $Fe(NO_3)_3/MoO_2$ catalyst, across trenches between prepatterned Pt source and drain electrodes[1] on $Si_3N_4/SiO_2$, as in our previous work[2]. The devices are probed, bonded and transferred to the cell within about an hour of removal from the furnace to minimize exposure to air. We did not find that annealing improved the isotherms. The trenches are 1 μm wide and 0.5 μm deep, with a Pt gate on the trench bottom. We select nanotubes with small gaps (20-100 meV) because they have better electrical contacts, and tend to be less noisy and more stable over long times, than those with large gaps.

*Apparatus*: The vapor cell is mounted in a closed-cycle cryostat with temperature range 4 – 300 K. The gas handling manifold and ceramic capacitance pressure gauge are at room temperature. The cell pressure $P$ is lower than the pressure $P_g$ read on an external capacitance gauge in static equilibrium due to thermal transpiration, and is obtained from it by applying a correction according to Ref. [3].

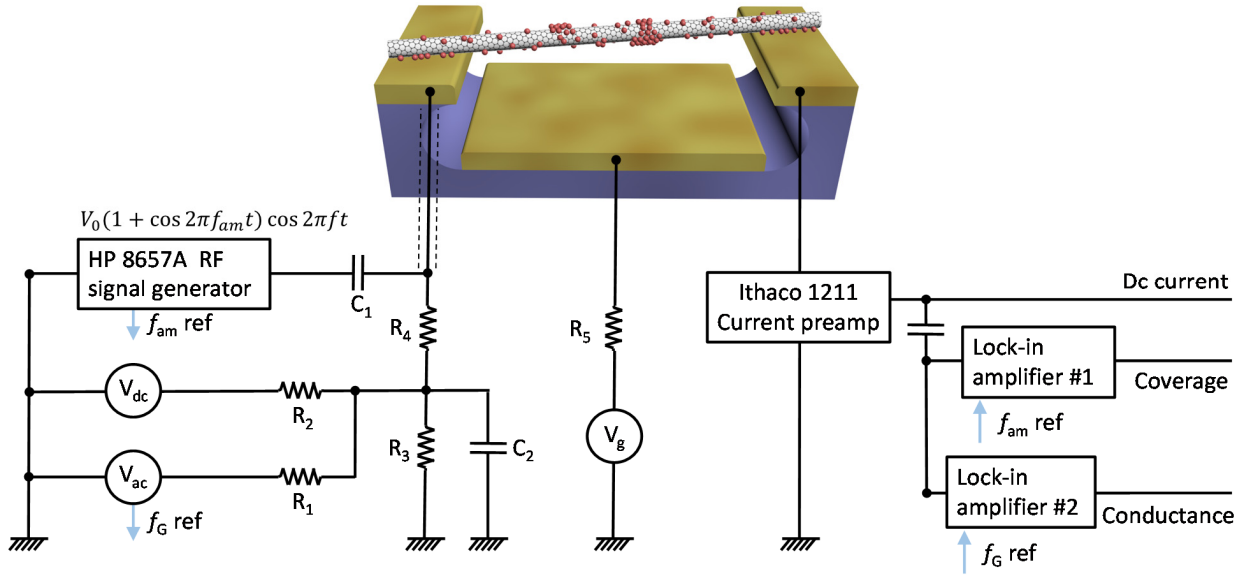

**Figure S1** | More detailed circuit for combined vibrational and conductance measurements. $R_1$, $R_2$ = 100 kΩ and $R_3$ = 1 kΩ form a low-frequency voltage divider/adder. $R_4$ = 10 kΩ and $C_2$ = 500 pF form a low-pass filter to block the RF signal from the ac and dc voltage source output circuits. $C_1$ = 500 pF removes a dc offset and low-frequency noise from the signal generator. $R_5$ = 10 MΩ protects the gate. A 50 Ω micro-coax takes the RF signal to the source contact. Other connections in the cryostat are simply copper wires twisted together.

*Conductance and coverage measurements*: Fig. S1 shows the electrical setup in more detail. The coverage $\phi$ is obtained from the shift of a mechanical resonance of the nanotube detected using a version of the mixing technique of Sazonova et al[4]. An RF signal at frequency $f$, amplitude-modulated at $f_{am}$ = 1 kHz and a depth of 100%, is applied only to the source of the nanotube device from a signal generator, and the resulting (mixing) current component $I_{mix}$ at $f_{am}$ is detected with a virtual-earth current preamplifier and a lock-in amplifier (lock-in #1). When $f$ is swept through a mechanical resonance of the nanotube (in the range 20-300 MHz) a feature is seen in $I_{mix}$ due to the increased nonlinearity of the conductance when the nanotube moves more. The effect of Ar pressure at 50 K is illustrated in Fig. S2 for device YB30. To measure the differential conductance we use an ac signal of amplitude $V_{ac}$ typically 1 mV and frequency $f_G \sim$ 200-600 Hz. This is combined with the RF signal and a dc bias for nonlinear measurements using the capacitors and resistors shown. The conductance is measured by lock-in #2 referenced to $f_G$. All signals, including pressure and temperature, are measured by a PC using LabView.

If the vibrations are not too nonlinear and the mass adsorbed is uniform, the resonance frequency shifts as $f \propto \text{mass}^{-1/2} \propto (m_C + \phi m_a)^{-1/2}$, from which the coverage is obtained as $\phi = N_a/N_C = (m_C/m_a)[(f_0/f_P)^2 - 1]$, where $f_0$ is the resonant frequency in vacuum and $f_P$ is the resonant frequency at an equilibrium gas pressure $P$, and $m_c$ and $m_a$ are the respective masses of the carbon atoms and the adsorbate atoms or molecules. Using this equation we find excellent consistency with, for example, the density at commensurate coverage[2] of Kr, and we obtain the same value of $\phi$ for different vibrational modes and different gate voltage.

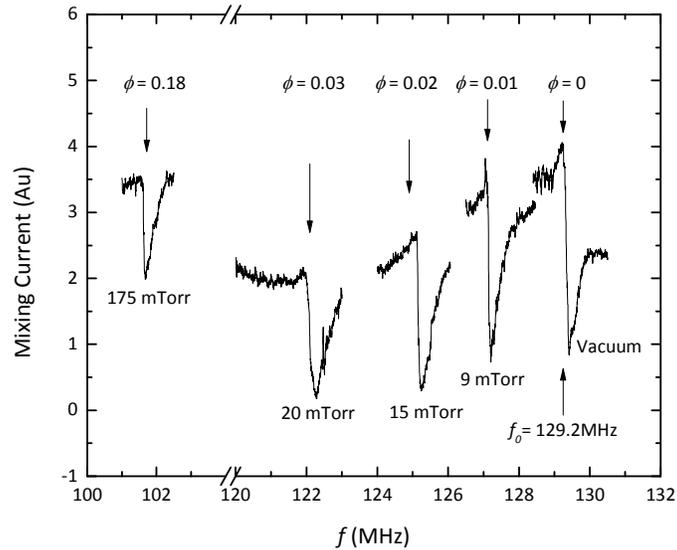

**Figure S2** | Example of the shift of a vibrational resonance in Ar vapor from which the coverage is derived. The shape of the resonance seen in the mixing current is dependent on gate voltage and drive amplitude, and importantly it broadens and distorts hugely in the steep regions of coverage isotherms. To obtain reliable coverage measurements in these regions we found that the RF drive amplitude $V_0$ should be no more than 1 mV. For many nanotubes the resonance could not be detected under these conditions.

## S2. Variation of the conductance sensitivity between devices and gases

Figures S3 and S4 show examples of the effect on the characteristics for different devices and different gases. The $G - V_g$ changes somewhat differently in each device, and in some cases $G$ increases while in others it decreases. However, the n- and p-thresholds, when they can be distinguished, shift in the same direction by roughly the same amount. Table 1 shows measured threshold shifts and corresponding charge transfers deduced from the period of Coulomb blockade oscillations seen at low temperatures.

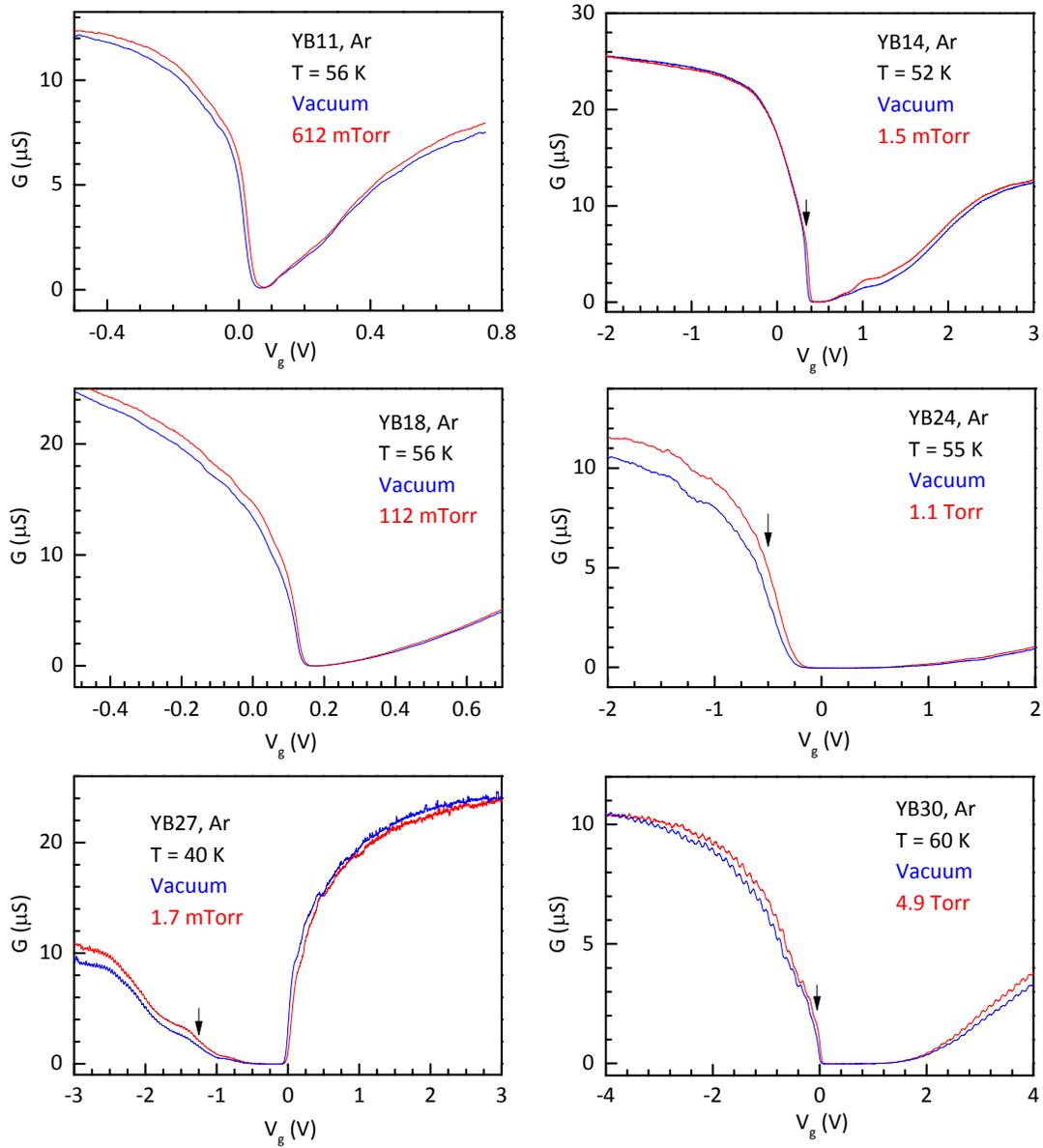

**Figure S3** | The effect of a dense monolayer of argon is shown on the characteristics of six suspended nanotube devices with similar geometry, including all those referred to in the main text, and one other device, YB18. The gate voltages used for conductance isotherms in the main text are indicated by arrows. In most devices the conductance is larger at negative than at positive $V_g$, implying lower Schottky barriers for *p*-channel than *n*-channel operation. However occasionally the opposite is the case (e.g. YB27 above).

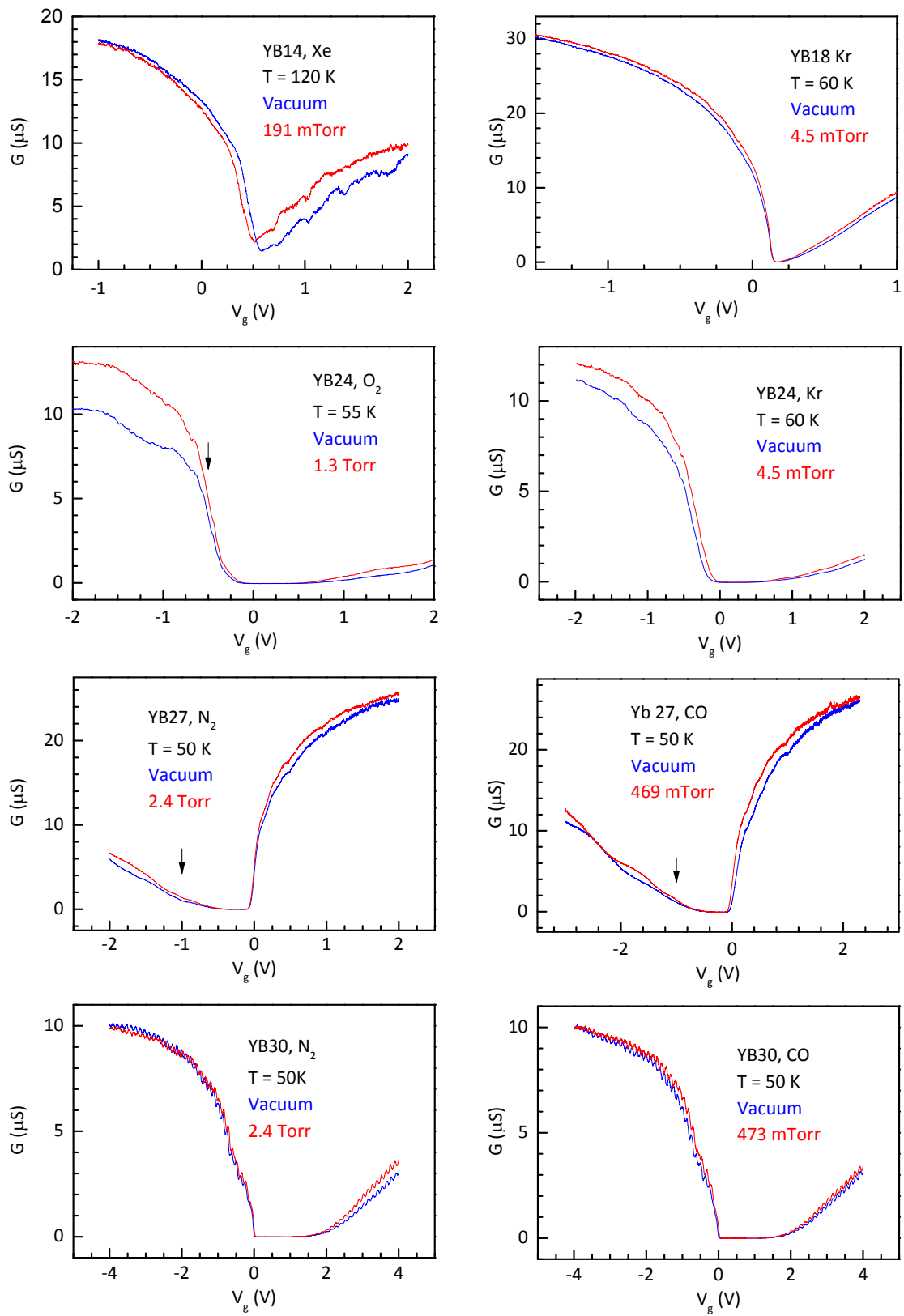

**Figure S4** | Examples of the effects of a dense monolayer of other gases, He, Kr, Xe, CO, $N_2$, and $O_2$, on the characteristics of several devices.

| Device YB# | CB spacing (mV) | Shift (mV) | Charge Transfer ($e$) |
|---|---|---|---|
| Ar | | | |
| 11 | 11 | +10 | +0.9 |
| 14 | 34 | +24 | +0.7 |
| 24 | 54 | +38 | +0.7 |
| 27 | 34 | +21 | +0.6 |
| 30 | 120 | +32 | +0.25 |
| | | | |
| Kr | | | |
| 14 | 34 | -12 | -0.35 |
| 24 | 54 | +69 | +1.3 |
| | | | |
| Xe | | | |
| 14 | 34 | -65 | -1.9 |
| | | | |
| He | | | |
| 14 | 34 | +23 | +0.7 |
| | | | |
| N$_2$ | | | |
| 14 | 34 | -13 | -0.4 |
| 27 | 34 | -7 | -0.2 |
| 30 | 120 | +20 | +0.2 |
| | | | |
| CO | | | |
| 27 | 34 | -35 | -1.0 |
| 30 | 120 | +30 | +0.25 |
| | | | |
| O$_2$ | | | |
| 24 | 54 | +27 | +0.5 |

Table S1. Gate voltage shifts and inferred total charge transfers for a dense monolayer ($\phi \approx 0.15$).

***Bound on the capacitance change***

We can put an upper bound on the change in capacitance to the gate caused by a monolayer of He. The gate voltage range spanned by 100 CB peaks changes by less than one peak spacing, implying a fractional capacitance change of $< 10^{-2}$. We can make a rough estimate of the expected change in capacitance taking a monolayer of He to be a 1 Å thick dielectric having the dielectric constant of liquid He, 1.055. Assuming a cylindrical capacitor geometry gives a fractional change in $C$ of $\frac{\Delta C}{C} \approx \frac{(1-1/\varepsilon)\Delta/a}{\ln b/a} < \sim 10^{-3}$, where $a = 1$ nm is the nanotube diameter, $b = 500$ nm is the distance from the nanotube to the gate electrode, and $\Delta = 0.1$ nm is the thickness of the monolayer.

*Possibility of contamination*

While some devices (such as YB14) reproduce the sharp phase transitions on clean graphite, indicating very high homogeneity, others (such as YB30) do not show vertical steps. It is natural to ask whether this observation, as well as the device-dependent sign of the conductance change at low coverage, can be explained by patches of contamination on the surface, such as adventitious carbon from the growth or organics deposited from the air. There are however a number of observations that are hard to reconcile with contamination. First, all devices show similar conductance responses to a dense monolayer, which implies that the response is predominantly of the same nature in this limit. Second, in all cases the frequency shift comes fairly close to the level expected for a monolayer coating of a clean nanotube, showing that the fraction of surface contaminated is small, since the mass of contaminants would reduce the frequency shift for a full monolayer. Third, the low-pressure parts of the coverage isotherms show no evidence of higher-binding inhomogeneities. Fourth, we have never seen intermediate cases that would correspond to a nanotube with some clean parts and some contaminated patches. Nevertheless we cannot rule out a role for contamination in some of the observations.

On the other hand, it is also possible that the variations are related mainly to the nanotube chirality/diameter via unanticipated mechanisms. Concerning the sign change, another possibility is that when the atoms cluster they polarize differently from when they do not. Future quantum chemistry calculations could investigate this possibility, which to our knowledge has not been considered before.

## S3. Phase diagram of Ar on graphite

In 1984 Migone, Li and Chan[5] constructed the phase diagram of monolayer Ar using heat capacity measurements (reproduced in Fig. S5). The fluid region and the melting line had been explored above 54 K by F. Millot[6] using volumetric adsorption isotherms. The incommensurate solid was studied with neutron diffraction by Taub et al.[7, 8], X-ray diffraction by J.P. McTague et al.[9] and K.L. D'Amico et al.[10], and using LEED by C.G. Shaw, S.C. Fain et al [11-13]. The monograph by Bruch, Zaremba and Cole[14] has a summary of the vapor pressure vs. temperature results and several additional references. A brief summary follows.

The density-temperature phase diagram in Fig. S5 resembles a simple van der Waals system, with a melting triple line at $\approx$ 47 K and liquid-vapor, solid-vapor and liquid-solid coexistence regions or boundaries. The L-V critical point is at $\approx$ 55 K. The flat shape of the L-V boundary coexistence region is consistent with the shape expected from the 2D Ising model, with critical exponent $\beta = 1/8$. Detailed thermodynamic and structural measurements though show that the system is quite complex, mostly due to the mismatch between the substrate structure and the Ar lattice in the solid or equilibrium distance between nearest neighbors in the liquid. The heat capacity measurements, run at approximately constant coverage, show that most of the entropy increase on going from the solid to the liquid below $n = 1$ (a commensurate density) does not occur at the triple line but broadly within the L-V coexistence region. Constant temperature LEED measurements at 42 K show that the uncompressed solid has a triangular structure with nearest-neighbor distance between Ar atoms of 0.394 nm and is rotated about 2.3° with respect to the orientation of a commensurate structure. The orientation angle increases with compression up to 3.7° at maximum compression, an effect predicted by A. D. Novaco and J. P. McTague[15, 16]. The constant coverage measurements of D'Amico et al. (done at $n = 0.81$, between 44.28 K and 50.03 K) give the same structure with nearest-neighbor distance of 0.392 nm and a solid lattice that expands through melting; the orientation angle decreases continuously as melting approaches, going from 2.9° at 44 K to about 2.2° at 49.5 K where the solid peak is gone. A small broad liquid peak remains oriented at 2° at 50 K, where the signal became difficult to measure.

Our isotherm data on device YB14 are consistent with the phase boundaries in Fig. S5, extend the isotherm results to below the 2D triple point, and yield for the first time the 2D compressibility critical exponent $\gamma$ at the L-V critical point.

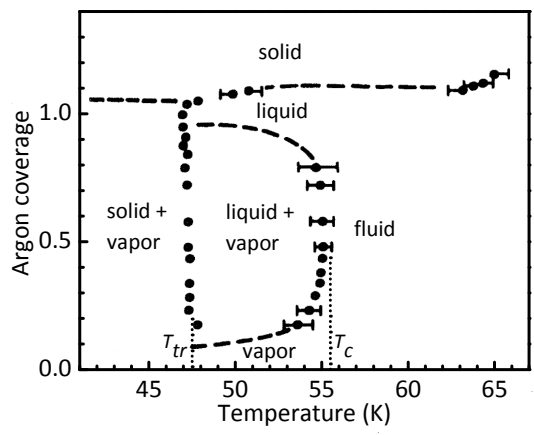

**Figure S5** | Coverage-temperature phase diagram for the first atomic layer of Ar on graphite. The data is from A. Migone et al.[5] (with permission). Here coverage = 1 corresponds to that of a commensurate monolayer, $\phi = 1/6$ in our terminology, where each argon atom sits on a 6-atom carbon ring.

**Supplementary References**